# A Case Study of Open Source Physics (OSP) Learning Community (LC)

*Lyna Kwan[1] & Loo Kang WEE[2]*

[1]*National Institute of Education,* Singapore

[2]Ministry of Education, Educational Technology Division, Singapore

[lyna@nie.edu.sg](mailto:lyna@nie.edu.sg), [Lawrence_WEE@moe.gov.sg](mailto:Lawrence_WEE@moe.gov.sg)

**Introduction**

Funded by the National Research Foundation (NRF) from 2010 to 2013, School of Science and Technology (SST) embarked on a project called WiMVT—Web-based iMVT. iMVT was meant to be an innovative pedagogy to learn science using modelling and visualization technology, and the web served as a platform to support students' collaborative inquiry-based learning. The project was one of the four proposals under the FutureSchools@Singapore (FS) Programme for SST. The word WiMVT was later changed into Collaborative Science Inquiry (CSI) after the new Principal Investigator (PI) took over the project as the original PI left the country. To spread out the good practices of the project to other schools in Singapore, Educational Technology Division (ETD) of Ministry of Education (MOE) attempts to formalize CSI Learning Community (LC), where Open Source Physics (OSP) LC is part of. Figure 1 depicts the CSI with the four key elements of it, i.e., Guided Inquiry, Modeling, Collaborative Discussion, and Visualization.

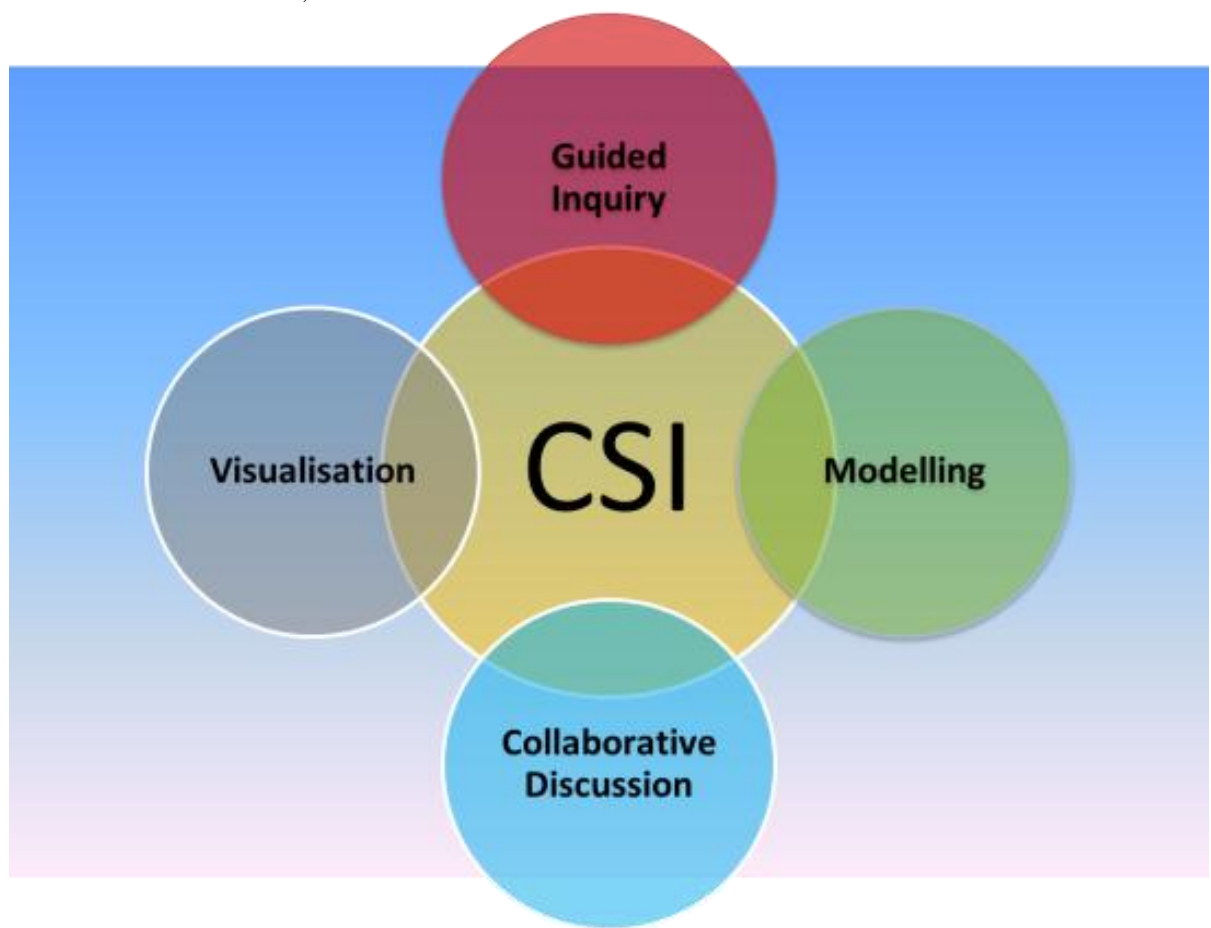

**Figure 1.** Collaborative Science Inquiry with the four key elements of it.

**What is Open Source Physics?**

OSP (Christian, Esquembre, & Barbato, 2011) is a National Science Foundation funded project that provides tools and resources for interactive computer-based modelling. Different from the common existing simulations and videos which are commonly used by teachers to teach science, computer-based modelling using OSP enables students and teachers to create, use, and customise the computer models or tools to suit their learning context and instructional needs (Wee & Mak, 2009), due to its open source codes.

Under the National Research Council's (NRC; 2012) *Framework for K-12 Science Education*, "modelling" and "mathematical and computational thinking" are listed among the scientific practices that are essential for learning science and engineering in grades K-12. These practices are also aligned with the MOE's goals for students engaging in critical and inventive thinking (MOE, 2015).

The OSP Collection of curriculum resources aims to engage students in physics through computer-based modelling. Computational physics and computer modelling provide new ways to understand, describe, explain, and predict scientific phenomena. The OSP Collection seeks to develop physics curriculum by using computational and modelling approaches which allow students to interact with the simulations and change the source codes that may lead to greater understanding and appreciation of scientific concepts. In short, combination of computational physics and computer modelling with theory and experiment allow deeper insight and understanding into the complexity of scientific phenomena that cannot be achieved with only one approach.

The key international community members of OSP in alphabetically order are:

1. Aaron Titus - High Point University
2. Anne Cox - Eckerd College
3. Bill Junkin - Eckerd College
4. Doug Brown - Cabrillo College
5. Francisco Esquembre - Universidad de Murcia
6. Harvey Gould - Clark University
7. Jan Tobochnik - Kalamazoo College
8. Mario Belloni - Davidson College
9. Wolfgang Christian - Davidson College

Not in this list, are a growing number of contributors that uses these tools to create many more curriculum materials all over the world, including Singapore's teachers.

**Open Source Physics@Singapore**
Below are the members of the Singapore OSP community. It is based on the Easy Java Script Simulation (EJSS) and Tracker lists maintained by the OSP LC facilitator, Wee Loo Kang. The combined number is about 23, of which the very active members are around 11. In alphabetically order,

1. Andy Luo Kangshun, Tampines Junior College
2. Chan Weng Cheong Ezzy, Peirce Secondary School
3. Chia Juin Wei, Curriculum Planning and Development Division (CPDD)-MOE
4. Dave Lommen, Hwa Chong Institution
5. Gideon Choo, River Valley High School
6. Goh Giam Hwee Jimmy, Yishun Junior College
7. Kwek Eng Yeow, Victoria Junior College
8. Lee Tat Leong, River Valley High, ETD-MOE

9. Leong Tze Kwang, Raffles Girls Secondary
10. Lim Ai Phing, River Valley High School
11. Lim Jit Ning, Hwa Chong Institution
12. Ling Yih Jye, Chung Cheng High (Yishun)
13. Lye Sze Yee, Teck Whye Primary School
14. Ng Boon Leong, Angelo-Chinese Junior College
15. Ng Kar Kit, River Valley High School
16. Ning Hwee Tiang, National Junior College
17. Samuel Ooi, National Junior College
18. Siow Seau Yan Sharon, Raffles Girls Secondary
19. Sng Peng Poo, Anderson Junior College
20. Tan Kim Kia, Evergreen Secondary School
21. Thio Cher Kuan, Raffles Girls Secondary
22. Wee Loo Kang Lawrence, ETD-MOE
23. Yeu Chee Wee Thomas, Meridian Junior College

Despite not having a formalized 23 core members-group of OSP LC so far, the 11 OSP activists occasionally meet, depending on opportunities, e.g., OSP workshops, seminars, conferences and MOE's Physics Instructional Programme Support Group (IPSG). There are three ways in which teachers appropriate OSP—create, adapt, and use.

**Creators for OSP@SG**
Out of the 23 members, there are at least three educators who *actively* create OSP simulations: Lim Jit Ning, who has many tracker video resources; Lee Tat Leong, who has many Easy Java Simulation (EJS) resources on his blog; and Wee Loo Kang Lawrence, who has many EJS and tracker video resources (see http://weelookang.blogspot.sg/p/physics-applets-virtual-lab.html). Some tracker videos contributed by the Singapore OSP community are available online at http://iwant2study.org/lookangejss/indexTRZdl_info.html

**Adapters of OSP@SG**
Due to reasons such as teachers' excessive workload (no time), teachers usually adapt the OSP worksheets more than adapt the OSP simulations although there are some teachers who occasionally adapt the simulations depending on the topics and their time. In general, there are a lot of teachers, local and overseas, who use some of resources created by our OSP@SG community. However, they do not inform us the details of the degree they use the resources, unless it is for new customization.

The growth of the OSP LC is difficult to measure since it is more like different agents advancing their own agenda on their own time and space while fulfilling the requirements of the job as a teacher in school. Some efforts have been made by Loo Kang to grow the number of possible creators, such as initiating three eduLab projects led by physics teachers entitled "Java Simulation Design for Teaching and Learning", "Becoming Scientists through Video Analysis" and Modeling and Interactive enabled Textbooks". So far, 4 experts from International OSP community were invited to Singapore to conduct a one to 4-day hands-on workshop, besides seminar, in order to empower teachers to create more resources.

**Users of OSP@SG**
While teachers can easily use existing OSP simulations, many of them encounter difficulties in adapting, let alone creating, OSP simulations. Some of the reasons are because not many of them have programming skills, it takes time to adapt or create OSP simulations, and so forth. The existing simulations such as PhET and NetLogo are other attractive alternatives that teachers can choose besides OSP, although the two simulation tools do not provide source codes which enable teachers to customize the simulations to address the developing needs of their students. Figure 2 illustrates the framework to build greater laterality in the OSP community in order to spread modelling practice using OSP.

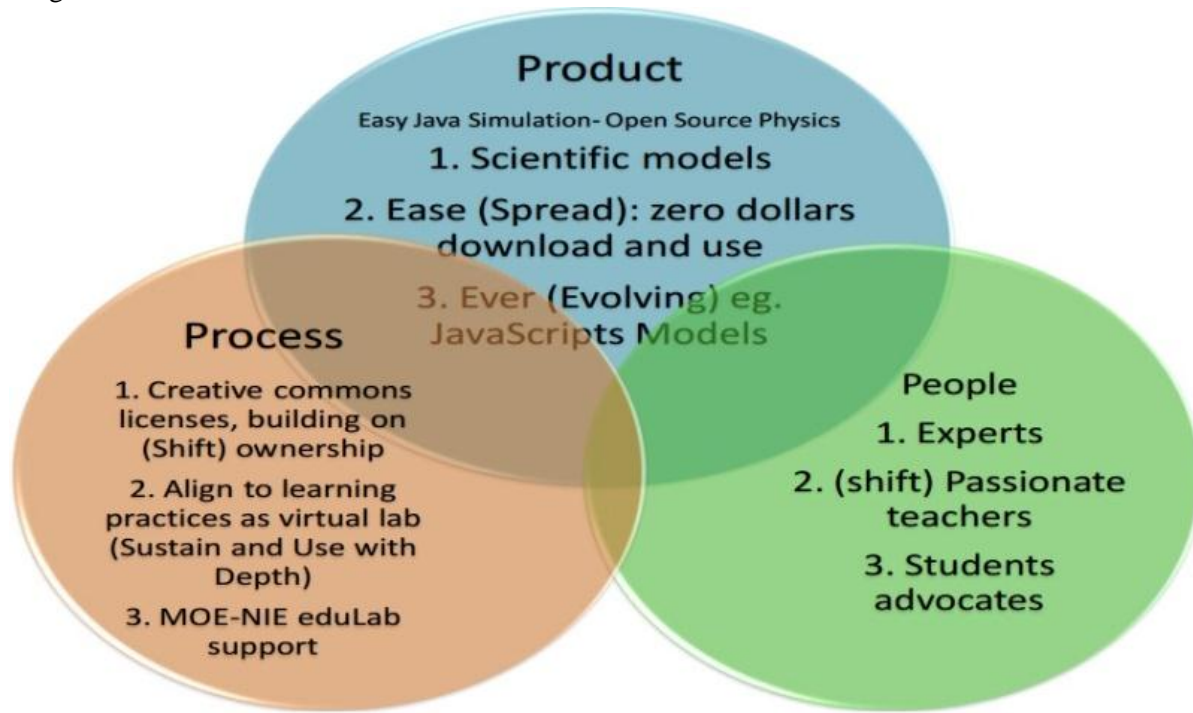

**Figure 2.** OSP 3Ps scaling-up framework with the dimensions of scaling up in brackets.
*[Adapted from Dede, C. (2007). Exploring the process of scaling up. Harvard University.]*

**OSP 3P's scaling-up framework**
- Product
  EJS does a great job to *produce* free download and use accurate scientific models that can be traced from the open source codes, remixed, and reused. Users do not need to login or server setup (Vargas et al., 2008). It only requires Java runtime (for EJS models) or a modern browser (for EJSS JavaScripts models).
- Process
  Aligning to existing *practice* of laboratory (Baser & Durmus, 2010; Dormido et al., 2008; Espinoza & Quarless, 2010; Jara et al., 2009), many of the EJS models can be used as a virtual laboratory to support experiential learning (Wee, 2012).
- People
  *Passionate educators* can adapt or create finer customized computer models to suit their technology, pedagogy, content, and context knowledge to better mould the learning experiences of their students. The key people in the OSP LC keep creating more computer models to suit their fellow teachers and students learning needs and release these computer models with activity worksheets and other resources for the benefit of all, which is the real motivation that drives our collaborative work using EJS.

**Vision, Reason, and Passion-Emotion – How OSP@SG was successful as LC?**
After all, it is vision (i.e., Open Source Physics Digital Library), reason (i.e., to benefit all learners), and passion (i.e., making Open Educational Resources as a "must" instead of a "should") that engineer and nurture the OSP LC. The clearer the vision, reason, and passion, the more enduring it is to overcome problems in scaling up the practices of the learning community. While there are some possibilities for students and teachers to gain deeper insights into scientific phenomena and deeper learning through computational modelling using OSP, strong support from educational systemic level will speed up and advance the agenda.